\date{}
\documentclass[aps,pra,showpacs]{revtex4-1}
\usepackage{graphicx,psfrag,amsmath,amssymb,amsfonts,latexsym,color,dcolumn}

\begin{document}

\title{Casimir Force for Absorbing Media in an Open Quantum System Framework: Scalar Model}

\author{Fernando C. Lombardo$^1$ \footnote{lombardo@df.uba.ar}}
\author{Francisco D. Mazzitelli$^{1,2}$ \footnote{fdmazzi@cab.cnea.gov.ar}}
\author{Adri\'an E. Rubio L\'opez $^1$\footnote{arubio@df.uba.ar}}

\affiliation{$^1$ Departamento de F\'\i sica {\it Juan Jos\'e
 Giambiagi}, FCEyN UBA, Facultad de Ciencias Exactas y Naturales,
 Ciudad Universitaria, Pabell\' on I, 1428 Buenos Aires, Argentina - IFIBA}
\affiliation{$^2$ Centro At\'omico Bariloche
Comisi\'on Nacional de Energ\'\i a At\'omica,
R8402AGP Bariloche, Argentina}

\date{today}

\begin{abstract}
In this article we compute the Casimir force between two finite-width mirrors at
finite temperature, working in a simplified model in $1+1$ dimensions. The mirrors, considered as dissipative media, are modeled
by a continuous set of harmonic oscillators which in turn are coupled to an external environment at
thermal equilibrium. The calculation of the  Casimir force is performed in the framework of the theory of
quantum open systems. It is shown that the Casimir interaction
has two different contributions: the usual radiation pressure from
vacuum, which is obtained for ideal mirrors without dissipation or losses, and a Langevin force
associated with the noise induced
by the interaction between dielectric atoms in the slabs and the thermal bath. Both contributions to
the Casimir force are needed in order to reproduce the analogous of Lifshitz
formula in $1+1$ dimensions. We also discuss the relation between the electromagnetic properties of the mirrors
and  the
spectral density of the environment.
\end{abstract}


\maketitle

\section{Introduction}

Given the precision that has been recently achieved in the measurement of the Casimir forces \cite{reviews},
the use of realistic models for the description of the media that
constitute the mirrors is an unavoidable step for the improvement of Casimir energy calculations,
which is needed for comparison with the experimental data. Moreover, from a conceptual point of view,
the theoretical calculations for mirrors with general electromagnetic properties, including
absorption,
is not a completely settled issue \cite{Barton2010,Philbin2010,daRosaetal}.  Since dissipative effects imply the
possibility of energy interchanges
between different parts of the full system (mirrors, vacuum field and environment), the theory of quantum open systems
\cite{Petruccione}
is the natural approach to clarify the role of dissipation
in Casimir physics. Indeed, in  this framework, dissipation and noise appears in the effective theory of the relevant
degrees of freedom (the electromagnetic field) after integration of the matter and other environmental degrees of freedom.

Dielectric slabs are in general nonlinear, inhomogeneous, dispersive and also dissipative media. These aspects
turn difficult the quantization of a field when all of them have to be taken into account simultaneously.
There are different approaches to address this problem. On the one hand, one can use a
phenomenological description based on
the macroscopic electromagnetic  properties of the materials. The quantization can be performed starting from
the macroscopic Maxwell equations,  and including noise terms to account for absorption
\cite{Buhmann2006}.
In this approach a canonical quantization scheme is not possible, unless one couples the electromagnetic field
to a reservoir (see \cite{Philbin2010}), following the standard route to include
dissipation in simple quantum mechanical systems. Another possibility is to establish a first-principles model
in which the slabs are described through their microscopic degrees of freedom, which are coupled
to the electromagnetic field. In this kind of models,  losses are also incorporated by considering a thermal bath, to allow for  the possibility of absorption of light.
There is a large body of literature on the quantization of the electromagnetic field in dielectrics.
Regarding microscopic models, the fully canonical quantization of the electromagnetic field
in dispersive and lossy dielectrics has been performed by Huttner and Barnett (HB) \cite{HB}. In the HB model,
the electromagnetic field
is coupled to matter (the polarization field), and the matter is coupled to a reservoir that is included into the model to
describe the losses. In the context of the theory of quantum open systems, one can think the HB model as a composite system
in which the relevant degrees of freedom belong to two subsystems (the electromagnetic field and the matter), and the matter degrees of freedom are in turn coupled to an environment (the thermal reservoir). The indirect
coupling between the electromagnetic field and the thermal reservoir is responsible for the losses. As we will comment below,
this will be our starting point
to compute the Casimir force between absorbing media.

Regarding the Casimir force, the celebrated Lifshitz formula \cite{Lifshitz} describes the forces between dielectrics
in terms of their macroscopic electromagnetic properties. The original derivation of this very general formula is based on a
macroscopic approach, starting from stochastic Maxwell equations and using thermodynamical properties
for the stochastic fields. As pointed out in several papers,  the connection between this approach and an
approach based on a fully quantized model is not completely clear.  Moreover,
some doubts have been  raised about the applicability of the Lifshitz formula to lossy dielectrics \cite{Barton2010,Philbin2010,daRosaetal}.

The first calculation of the Casimir force between two absorbing slabs using a microscopic approach is,
to our knowledge,  due to Kupiszewska \cite{Dorota},
 who modeled dielectric atoms as a set of harmonic oscillators coupled to  an environment
 with $T=0$ , in which the atoms can
dissipate energy. In that work, a scalar field in 1+1 dimensions was considered, and all the environmental effects
were described through a dissipative constant and a Langevin force.
In the context of quantum open systems, this is tantamount to consider an Ohmic environment.
The force between slabs was then obtained in terms of the reflection coefficients associated to the slabs, which are described by
a complex dielectric function. This result was rederived using a Green-function method for quantizing
the macroscopic field in absorbing systems in 1D in conjuction with scattering matrix approach \cite{matloob}. This was also
extended to two identical absorbing superlattices \cite{esquivel1}. Esquivel-Sirvent et al demonstrated an alternative
Green-function approach that makes the quantization of the field within the slabs unnecessary and calculate the Casimir
force in an asymmetric configuration \cite{esquivel2} which was earlier considered only in the lossless case \cite{jaekel}. In Ref.\cite{tomas},
the Casimir force was calculated in a lossless dispersive layer of an otherwise absorbing multilayer by using the macroscopic field operators
considered by \cite{gruner}. In  the series of works \cite{daRosaetal,daRosaetal2,daRosaetal3}, Rosa et al considered the 
evaluation of the electromagnetic energy density in the presence of an absorbing and dissipative dielectric, using the HB model
for $T=0$ and constant dissipation. In particular, in the recent  Ref.\cite{daRosaetal3}, they obtained the force density associated to
spatial variations of the permittivity from which, in principle, one could obtain the Casimir force between slabs. 

In this paper we will follow a program similar to that of Ref.\cite{Dorota}, generalizing it by considering a general and
well defined quantum open system. We will work with a simplified model analogous to the one of HB,   assuming that the
dielectric atoms in the slabs are quantum Brownian particles,
and that they are subjected to fluctuations (noise) and dissipation, due to the coupling to an external thermal environment. We will keep
generality in the type of spectral density to specify the
bath to which the atoms are coupled, generalizing the constant dissipation
model used in Ref.\cite{Dorota}. Indeed,  after integration of the environmental degrees of freedom, it will be possible to obtain
the dissipation and noise kernels that modify the unitary equation of motion of the dielectric atoms.
As we will see,  general non-Ohmic environments do not provide constant
dissipation coefficients in the equation of motion of the Brownian particles, even at high temperatures.  Moreover, the spectral density of the environment determines the electromagnetic properties of the mirrors and therefore have a direct influence
on the Casimir force.

In addition to the conceptual issues described above, there are additional motivations to consider detailed microscopic models
of the Casimir force, in particular, the controversy about its temperature dependence. Assuming 
simple phenomenological descriptions of the materials based on  the plasma or Drude models, the
theoretical  predictions
for the Casimir force are different due to the contribution (or not) of the TE zero mode  \cite{Brevik}.
At small distances $a$ such that $aT\ll 1$, the differences are not too large, and the experimental results by Decca et al 
\cite{Decca} seem to be
well described by the plasma model. However, at large distances $aT\gg 1$, the theoretical predictions differ by a factor $2$.
The Casimir force at such large distances have been recently measured \cite{naturephys}, and the results are compatible with the Drude 
model after taking into account  the interactions due to patch potentials on the surfaces of the conductors
(some authors disagree with the evaluation of the effects of the patch potentials, see  \cite{crit}). In any case, these controversies 
show that more detailed microscopic models are necessary to clarify the situation.  For example, considering 
that the slabs contain 
classical or quantum non relativistic charges interacting via the static Coulomb potential,  the result for the 
large distance limit agrees with that of the Drude model \cite{BM}.

Another motivation for considering the Casimir forces in the framework of open quantum systems 
is the possibility of analyzing non equilibrium effects, such as the Casimir force between
objects at different temperatures \cite{Antezza} and the power of heat transfer
between them \cite{Bimonte}, including the time dependent evolution until reaching a stationary situation.

The paper is organized as follow: in next Section we shall present the model, the  Heisenberg equations of motion
for the different operators, and the
vacuum and Langevin contributions to the field operator.
In Section III we study much in deep the relationship between
the microscopic model and the macroscopic electromagnetic properties of the mirrors.
Section IV is
dedicated to the evaluation of the Casimir force. After adding the vacuum and Langevin contributions, we show
that the total force is given by a Lifshitz-like formula, where the reflection coefficients of the slabs depend on
the properties of the atoms and the environment considered in the model.   In Section V we comment on
the relation of the quantum open systems approach developed in this paper and the Euclidean computation of
the Casimir force. We summarize our findings
in Section VI. The Appendices contain some details of the calculations.

\section{The Model}

\subsection{The Lagrangian Density}

With the aim of including effects of dissipation and noise in the calculation of
Casimir force, we will use the theory of open quantum systems, having in mind the paradigmatic example
of the quantum Brownian motion (QBM) \cite{Petruccione}.

The model consists of a system composed of two parts: a  massless scalar field and dielectric slabs which, in turn, are described by their internal degrees of freedom (a set of harmonic oscillators). Both sub-systems conform a composite system which is coupled to a second  set of harmonic oscillators, that plays the role of an external environment or thermal bath.
For simplicity we will work in $1+1$ dimensions. In our toy model
the massless field represents the electromagnetic field, and the first set of harmonic oscillators directly coupled to the
scalar field represents the atoms in the slab.

Considering the usual interaction term between the electromagnetic field and the ordinary matter, the coupling between the field and the atoms in the slab will be taken as a current-type one, where the field couples to the velocity of the atoms.
The coupling constant for this interaction is the electric charge $e$. We will also assume that there is no direct coupling between the field and the thermal bath. The Lagrangian density is therefore given by:

\begin{eqnarray}
\cal{L}&=&\cal{L}_{\phi}+\cal{L}_{S}+\cal{L}_{\phi-S}+\cal{L}_{B}+\cal{L}_{S-B}\nonumber\\&=&\frac{1}{2}
\partial_{\mu}\phi\partial^{\mu}\phi+4\pi\eta\left(\frac{1}{2}m\dot{r}^{2}(x;t)-\frac{1}{2}m\omega_{0}^{2}r^{2}(x;t)\right)+4\pi\eta
e\phi(x;t)\dot{r}(x;t)\nonumber\\&+&4\pi\eta\sum_{n}\left(\frac{1}{2}m_{n}\dot{q}_{n}^{2}(x;t)-\frac{1}{2}m_{n}\omega_{n}^{2}q_{n}^{2}(x;t)\right)-4\pi\eta\sum_{n}\lambda_{n}q_{n}(x;t)r(x;t),
\label{LagrangianaTOTAL}
\end{eqnarray}
where we have stressed the fact that $r$ and $q_{n}$ are also functions of the position, i.e.  each atom interacts with a thermal bath placed at the same position.
We have denoted by $\eta$ the density of atoms in both slabs.
The constants $\lambda_{n}$ are the coupling constants between the atoms and the bath oscillators.
It is implicitly understood that Eq.(\ref{LagrangianaTOTAL}) represents the Lagrangian density inside the plates, while outside
the plates
the Lagrangian is given by the free field one. The configuration of the slabs,  of thickness $d$ and separated by a distance
$a$ is shown in Fig.1.

\begin{figure}
\centering
\includegraphics[width=8cm , angle=0]{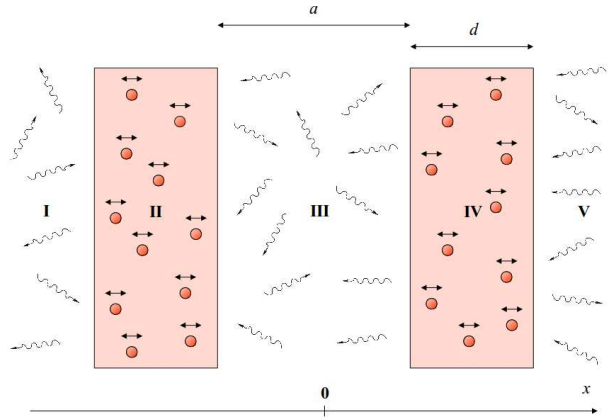}
\caption{(Color online). Plates configuration defining the five regions in space. The plates are formed by oscillators which are in contact with thermal environments. All the system is in thermal equilibrium at a temperature $T$. In this work we consider a 1+1 version of this configuration.} \label{FiguraUnica}
\end{figure}

The quantization of the theory is straightforward.  It should be noted that the full Hilbert space of the model $H$, where the quantization is performed, is not only the field Hilbert space $H_{\phi}$ (as is considered in others works where the field is the only relevant degree of freedom), but also includes the Hilbert spaces of the atoms $H_{S}$ and the bath oscillators $H_{B}$, in such a way that $H=H_{\phi}\otimes H_{S}\otimes H_{B}$. We will assume, as frequently done in the context of QBM,  that for $t<0$ the three parts of the systems are uncorrelated and not interacting. Interactions are turned on at $t=0$. Therefore,
the initial conditions for the operators $\widehat{\phi}$, $\widehat{r}$ must be given in terms of operators acting in each part of the Hilbert space. The interactions will make that initial operators to  become operators over the whole space $H$.
The initial density matrix of the total system is of the form:

\begin{equation}
\widehat{\rho}(0)=\widehat{\rho}_{\phi}(0)\otimes\widehat{\rho}_{S}(0)\otimes\widehat{\rho}_{B}.
\label{EstadoInicial}
\end{equation}
Since we are interested in the steady state ($t\rightarrow +\infty$), we will assume thermal equilibrium at temperature $T=1/\beta$ between the three parts. Each density matrix in Eq.(\ref{EstadoInicial}) will be taken as thermal type (we have set $\hbar = k_B = c = 1$).

\subsection{Heisenberg equations of motion}

Starting from the Lagrangian (\ref{LagrangianaTOTAL}), it is easy to derive the Heisenberg equations of motion for the
different operators. They are are given by:

\begin{equation}
\widehat{p}_{n}=m_{n}\dot{\widehat{q}}_{n},
\label{EcMovTOTALA}
\end{equation}

\begin{equation}
\dot{\widehat{p}}_{n}=-m_{n}\omega_{n}^{2}\widehat{q}_{n}+\lambda_{n}\widehat{r},
\label{EcMovTOTALB}
\end{equation}

\begin{equation}
\widehat{p}=m\dot{\widehat{r}}+e\widehat{\phi},
\label{EcMovTOTALC}
\end{equation}

\begin{equation}
\dot{\widehat{p}}=-m\omega_{0}^{2}\widehat{r}+\sum_{n}\lambda_{n}\widehat{q}_{n},
\label{EcMovTOTALD}
\end{equation}

\begin{equation}
\square\widehat{\phi}=4\pi\eta e\dot{\widehat{r}},
\label{EcMovTOTALE}
\end{equation}
where the operators $\widehat{p}$ and $\widehat{p}_{n}$ are the conjugate momentum operators associated to the operators $\widehat{r}$ and $\widehat{q}_{n}$ respectively.
Substituting Eq.(\ref{EcMovTOTALA}) into Eq.(\ref{EcMovTOTALB}), we get:

\begin{equation}
m_{n}\ddot{\widehat{q}}_{n}+m_{n}\omega_{n}^{2}\widehat{q}_{n}-\lambda_{n}\widehat{r}=0.
\label{EcMovQN}
\end{equation}

As usual in the context of QBM,  we solve the equations for the operators $\widehat{q}_{n}$ taking $\widehat{r}$ as a source, and replace  the solutions into Eq.(\ref{EcMovTOTALD}). In this way, the microscopic degrees of freedom
in the slabs satisfy a Langevin-like equation of the form

\begin{equation}
\dot{\widehat{p}}=-m\omega_{0}^{2}\widehat{r}-m\frac{d}{dt}\int_{0}^{t}d\tau\gamma(t-\tau)\widehat{r}(x;\tau)+\widehat{F}(x;t),
\label{EcMovRYP}
\end{equation}
where the damping kernel $\gamma$ and the stochastic force operator $\widehat{F}$ are the same as the ones of QBM (see \cite{Petruccione} for a general and complete view). They are given by

\begin{equation}
\gamma(t)=\frac{2}{m}\int_{0}^{+\infty}d\omega\frac{J(\omega)}{\omega}\cos(\omega t),
\label{QBMNucleoAmortiguamiento}
\end{equation}

\begin{equation}
\widehat{F}(t)=\sum_{n}\frac{\lambda_{n}}{\sqrt{2m_{n}\omega_{n}}}\left(e^{-i\omega_{n}t}\widehat{b}_{n}+e^{i\omega_{n}t}\widehat{b}_{n}^{\dag}\right).
\label{QBMOperadorFdet}
\end{equation}
Here $\widehat{b}_{n}$ and $\widehat{b}_{n}^{\dag}$ are the annihilation and creation operators associated
to $\widehat{q}_{n}$, and $J(\omega)$ is the spectral density that characterizes the environment. This function gives the number of oscillators in each frecuency for given values of the coupling constants $\lambda_{n}$:

\begin{equation}
J(\omega)=\sum_{n}\frac{\lambda_{n}^{2}}{2m_{n}\omega_{n}}\delta\left(\omega-\omega_{n}\right).
\label{QBMDensidadEspectral}
\end{equation}

In order to obtain a true irreversible dynamics, we introduce a continuous distribution of bath modes, that replaces the spectral density by a smooth function of the frequency $\omega$ of the bath modes. Different functions will describe different types of environments. Physically, the thermal bath has a finite number of oscillators in a given range of frequencies. Then, a cutoff function must be introduced, containing some characteristic frecuency scale $\Lambda$. Then, the spectral density takes the following form

\begin{equation}
J(\omega)=\frac{2}{\pi}m\gamma_{0}\omega\left(\frac{\omega}{\Lambda}\right)^{\alpha-1}\textit{f}\left(\frac{\omega}{\Lambda}\right).
\label{QBMDensidadEspectralEntornos}
\end{equation}
In the equation above, $\gamma_{0}$ is the relaxation constant of the environment, while $f$ is the frequency cutoff function.
The values of $\alpha$ classify the different types of environments: $\alpha=1$ corresponds to an ohmic environment (in which there is a dissipative term proportional to the velocity in the Langevin equation for the Brownian particle),  while $\alpha<1$ and
$\alpha>1$ describe subohmics and supraohmics environments respectively \cite{Petruccione}.

%%%%%%%

At equilibrium, the stochastic force operator in Eq.(\ref{QBMOperadorFdet}) and the damping kernel $\gamma$ in Eq.(\ref{QBMNucleoAmortiguamiento}) are not independent. The statistical properties of the stochastic force operator are given by the dissipation and noise kernels

\begin{equation}
D(t-t')\equiv
i\left\langle\left[\widehat{F}(t);\widehat{F}(t')\right]\right\rangle=i\left[\widehat{F}(t);\widehat{F}(t')\right]=-2\int_{0}^{+\infty}d\omega J(\omega)\sin[\omega (t - t')],
\label{QBMNucleoDisipacion}
\end{equation}

\begin{equation}
D_{1}(t-t')\equiv
\left\langle\left\{\widehat{F}(t);\widehat{F}(t')\right\}\right\rangle=2\int_{0}^{+\infty}d\omega J(\omega)\coth\left(\frac{\omega}{2T}\right)\cos[\omega (t - t')],
\label{QBMNucleoRuido}
\end{equation}
which are the formal quantum open systems generalization of the relations employed in Ref.\cite{Dorota} for general
 environments and arbitrary temperature. Note that only the noise kernel $D_{1}$ involves the environmental temperature $T$ as a parameter. Considering Eq.(\ref{QBMNucleoAmortiguamiento}) and Eq.(\ref{QBMNucleoDisipacion}), it is easy to show that:

\begin{equation}
\frac{d}{dt}\gamma(t-s)=-\frac{1}{m}D(t-s),
\label{QBMNucleoAmortiguamientoProp1}
\end{equation}
which relates the damping kernel $\gamma$ to the statistical properties of the stochastic force operator $\widehat{F}$.

All in all, the set of equations to solve now are Eqs.(\ref{EcMovTOTALC}), (\ref{EcMovTOTALE}), and (\ref{EcMovRYP}).

It is possible to obtain a formal solution for the operators $\widehat{r}(x;t)$
by  considering the field $\phi$ as a source for the equation. This solution generalizes the crude approximation made in Ref. \cite{Dorota} for the evolution of the microscopic degrees of freedom in the mirrors. It is given by:

\begin{equation}
\widehat{r}(x;t)=G_{1}(t)\widehat{r}(x;0)+G_{2}(t)\dot{\widehat{r}}(x;0)+\frac{1}{m}\int_{0}^{t}dsG_{2}(t-s)\left(\widehat{F}(x;s)-e\dot{\widehat{\phi}}(x;s)\right),
\label{SolucionQBMAcoplado}
\end{equation}
where $G_{1,2}$ are the Green functions associated to the QBM equation that satisfy:

\begin{align}
G_{1}(0)=1\;\;\;\;\text{,}\;\;\;\;\dot{G}_{1}(0)=0,
\label{CondInicialesG1}
\end{align}

\begin{align}
G_{2}(0)=0\;\;\;\;\text{,}\;\;\;\;\dot{G}_{2}(0)=1,
\label{CondInicialesG11}
\end{align}
for which, the Laplace transforms are given by

\begin{equation}
\widetilde{G}_{1}(z)=\frac{z}{z^2+\omega_{0}^2+z\widetilde{\gamma}(z)},
\label{TransformadaG1}
\end{equation}

\begin{equation}
\widetilde{G}_{2}(z)=\frac{1}{z^2+\omega_{0}^2+z\widetilde{\gamma}(z)},
\label{TransformadaG2}
\end{equation}
where $\widetilde{\gamma}$ is the Laplace transform of the damping kernel.
Note that,  given these conditions, one can  prove that $G_{1}(t)=\dot{G}_{2}(t)$.

Inserting this solution into Eq.(\ref{EcMovTOTALE}), we obtain the following equation for the field operator

\begin{equation}
\square\widehat{\phi}+\frac{4\pi\eta
e^2}{m}\int_{0}^{t}{G}_{1}(t-\tau)\dot{\widehat{\phi}}(x;\tau)d\tau = 4\pi\eta
e\left(\dot{G}_{1}(t)\widehat{r}(x;0)+G_{1}(t)\dot{\widehat{r}}(x;0) +\frac{1}{m}\int_{0}^{t}{G}_{1}(t-\tau)\widehat{F}(x;\tau)d\tau\right),
\label{EcMovCampoTOTAL}
\end{equation}
subjected to the free field initial conditions:

\begin{equation}
\widehat{\phi}(x;0)=\int dk\left(\frac{1}{\omega_{k}}\right)^{\frac{1}{2}}\left(\widehat{a}_{k}e^{ik x}+\widehat{a}_{k}^{\dag}e^{-ik x}\right),
\label{CondIniCampoTOTAL}
\end{equation}

\begin{equation}
\dot{\widehat{\phi}}(x;0)=\int dk\left(\frac{1}{\omega_{k}}\right)^{\frac{1}{2}}\left(-i\omega_{k}\widehat{a}_{k}e^{ik x}+i\omega_{k}\widehat{a}_{k}^{\dag}e^{-ik x}\right),
\label{CondIniDerCampoTOTAL}
\end{equation}
where $\widehat{a}_{k}$ and $\widehat{a}_{k}^{\dag}$ are the annihilation and creation operators for the free field, and $\omega_{k}=|k|$. The  boundary conditions are the continuity of the field and its spatial derivative at the interface points.

We will compute the Casimir force from the $xx$-component of the energy-momentum tensor
\begin{equation}
\widehat{T}_{xx}(x;t)=\frac{1}{2}\left((\partial_{0}\widehat{\phi})^2+(\partial_{x}\widehat{\phi})^2\right).
\label{T001D}
\end{equation}
The force is explicitly given by
\begin{equation}
F_{\rm C}=\langle \widehat{T}_{xx}^{\rm ext}\rangle-\langle
\widehat{T}_{xx}^{\rm int}\rangle,  \label{FCasimirPresion}
\end{equation}
where the expectation values are taken on the regions outside the planes (Regions I or V) and between them (Region III), respectively,  in a thermal equilibrium situation.

For this calculation, we need the explicit solution for the field equation
(\ref{EcMovCampoTOTAL}).  Considering the properties of the fundamental solutions $G_{1,2}$, the first step is to take the Laplace transform of the equation in order
to obtain

\begin{eqnarray}
\frac{\partial^{2}}{\partial
x^2}\widehat{\phi}(x;s)&-& s^2\left(1+\frac{4\pi\eta
e^{2}}{m}\widetilde{G}_{2}(s)\right)\widehat{\phi}(x;s)+s\widehat{\phi}(x;0)+\dot{\widehat{\phi}}(x;0)+\frac{4\pi\eta
e}{m}\widetilde{G}_{1}(s)\widehat{F}(x;s)+\frac{4\pi\eta
e}{m}\widetilde{G}_{1}(s)\widehat{p}(x;0)\nonumber\\&&+4\pi\eta
e\widehat{r}(x;0)\left(s\widetilde{G}_{1}(s)-1\right)=0.
\label{EcMovCampoTransformada1DTOTALConR0yP0}
\end{eqnarray}

Since we are interested in  the long-time behaviour ($t\rightarrow+\infty$), we can omit terms containing the positions $\widehat{r}$ and momenta $\widehat{p}$ of the oscillators at $t=0$. This assumption is well justified for  $t\gg1/\gamma_0$,  the scale associated with the damping or relaxation time of the environment. Therefore, the equation for the field can be approximated by:

\begin{equation}
\frac{\partial^{2}}{\partial
x^2}\widehat{\phi}(x;s)-s^2\left(1+\frac{4\pi\eta
e^{2}}{m}\widetilde{G}_{2}(s)\right)\widehat{\phi}(x;s)=-s\widehat{\phi}(x;0)-\dot{\widehat{\phi}}(x;0)-\frac{4\pi\eta
e}{m}\widetilde{G}_{1}(s)\widehat{F}(x;s).
\label{EcMovCampoTransformada1DTOTAL}
\end{equation}
This is the  equation for the Laplace transform of the field, with the initial conditions and the Laplace transform of the stochastic Langevin force as sources. For simplicity, the spatial dependence of the matter terms was omitted, but it is important to remember that this expression is valid for the points inside the plates.
Note also that, as there is no contribution from the atoms to the sources,
the field operator $\widehat{\phi}$ acts on $H_{S}$ as an identity.

We propose a solution of the form $\widehat{\phi}(x;t)=\widehat{\phi}_{V}(x;t)+\widehat{\phi}_{L}(x;t)$, where  two contributions are distinguished: the vacuum contribution $\widehat{\phi}_{V}(x;t)$, which results from the modified field modes; and the Langevin contribution $\widehat{\phi}_{L}(x;t)$, which depends linearly on the Langevin forces. Each contribution satisfies 

\begin{equation}
\frac{\partial^{2}}{\partial x^2}\widehat{\phi}_{V}(x;s)-s^2\left(1+\frac{4\pi\eta e^{2}}{m}\widetilde{G}_{2}(s)\right)\widehat{\phi}_{V}(x;s)=-s\widehat{\phi}(x;0)-\dot{\widehat{\phi}}(x;0),
\label{EcMovCampoTransformadaVacio1DTOTAL}
\end{equation}

\begin{equation}
\frac{\partial^{2}}{\partial x^2}\widehat{\phi}_{L}(x;s)-s^2\left(1+\frac{4\pi\eta e^{2}}{m}\widetilde{G}_{2}(s)\right)\widehat{\phi}_{L}(x;s)=-\frac{4\pi\eta e}{m}\widetilde{G}_{1}(s)\widehat{F}(x;s).
\label{EcMovCampoTransformadaLangevin1DTOTAL}
\end{equation}

It is worth to note that the first equation presents only operators acting on $H_{\phi}$, thus the associated field contribution is an operator on that space. In the same way, the second equation depends only on operators acting on $H_{B}$,
and therefore the Langevin contribution acts nontrivially only there. Taking all this into account
the field operator reads
\begin{equation}
\widehat{\phi}(x;t)=\widehat{\phi}_{V}(x;t)\otimes\mathbb{I}_{S}\otimes\mathbb{I}_{B}+\mathbb{I}_{\phi}\otimes\mathbb{I}_{S}\otimes\widehat{\phi}_{L}(x;t).
\label{SeparacionContribuciones}
\end{equation}

In summary, the atoms act like a {\it bridge} between the field and the thermal bath, making no contributions to the total field. The problem, then, has been completely separated into two parts, that will be computed in the next subsection.

\subsection{Vacuum and Langevin contributions}

We will now solve the two independent Eqs. (\ref{EcMovCampoTransformadaVacio1DTOTAL}) and (\ref{EcMovCampoTransformadaLangevin1DTOTAL}). For the vacuum contribution $\widehat{\phi}_{V}$, the solution
for $t\rightarrow+\infty$ is assumed to have the form:

\begin{equation}
\widehat{\phi}_V(x;t)=\left[\int_{0}^{+\infty}+\int_{-\infty}^{0}\right]\frac{dk}{2\pi}\left(\frac{\pi}{\omega_{k}}\right)^{\frac{1}{2}}\left(\widehat{a}_{k}e^{-i\omega_{k}t}f_{k}(x)+\widehat{a}_{k}^{\dag}e^{i\omega_{k}t}f_{k}^{*}(x)\right),
\label{OperadorPhiDielectrico}
\end{equation}
where the first integral comprises the waves going from left to right, and the second one the waves going from right to left. The mode functions $f_{k}(x)$ satisfy the equation:

\begin{equation}
\frac{d^2}{dx^2}f_{k}(x)+\omega_{k}^{2}n^2(\omega_{k})f_{k}(x)=0.
\label{EcModosVacio1DTOTAL}
\end{equation}
The refractive index $n(\omega_{k})$ given by

\begin{equation}
n^2(\omega_{k})=1+\frac{4\pi\eta e^2}{m}\widetilde{G}_{2}(-i\omega_{k})=1+\frac{\omega_{P}^{2}}{\omega_{0}^{2}-\omega_{k}^{2}-i\omega_{k}\widetilde{\gamma}(-i\omega_{k})},
\label{IndiceRefraccionTOTAL}
\end{equation}
where $\omega_{\rm P}^{2}=\frac{4\pi\eta e^2}{m}$ is the plasma frecuency.

It is worthy to note that Eq.(\ref{EcModosVacio1DTOTAL}) is of the same form that the equation for the modes in a nonabsorbing dielectric medium \cite{Dorota90}, except that in this case the refractive index is frecuency-dependent. On the other hand,  Eq.(\ref{IndiceRefraccionTOTAL}) depends on the Fourier transform of the damping kernel, which is associated to a general environment. There is no assumption neither about the spectral density of the environment nor on the value of the equilibrium temperature. In the next Section we will analyze in more detail the relation between the electromagnetic response of the medium
and the spectral density.

The Eq. (\ref{EcModosVacio1DTOTAL}) for the modes can be solved by imposing continuity of the mode functions $f_k$ at all interfaces. The calculation is long but straightforward and is described in Appendix A.

In order to obtain the Langevin contribution, we must solve Eq. (\ref{EcMovCampoTransformadaLangevin1DTOTAL}). Since the interaction begins at $t=0$, the equations for the Laplace and Fourier transforms for this contribution are identical, and related by $s=-ik$. Thus,  the long-time solution can be written as $\widehat{\phi}_{L}(x;t)=\int_{-\infty}^{+\infty}\frac{dk}{2\pi}\widehat{\phi}_{L}(x;k)e^{-ikt}$, where $\widehat{\phi}_{L}(x;k)$ satisfies

\begin{equation}
\frac{\partial^{2}}{\partial x^2}\widehat{\phi}_{L}(x;k)+k^2\widehat{\phi}_{L}(x;k)=0,
\label{EcMovCampoFourierLangevin1DReg135TOTAL}
\end{equation}
in regions I, III and V; and,

\begin{equation}
\frac{\partial^{2}}{\partial x^2}\widehat{\phi}_{L}(x;k)+k^{2}n^{2}(k)\widehat{\phi}_{L}(x;k)=-\frac{4\pi\eta e}{m}\frac{ik}{\left(k_{0}^{2}-k^{2}-ik\widetilde{\gamma}(-ik)\right)}\widehat{F}(x;k),
\label{EcMovCampoFourierLangevin1DReg24TOTAL}
\end{equation}
in regions II and IV. Here  $\widehat{F}(x;k)$ is the Fourier transform of the stochastic force operator.  The explicit solution
is presented in Appendix B.

In Section IV we will use the vacuum and Langevin contributions to the field operator in order to obtain the Casimir force between slabs. Before doing that, we will describe in more detail the relation between the macroscopic electromagnetic properties of
the slabs and the microscopic model.

\section{Generalized Permittivity from the quantum open system}

With the aim of checking if our model is physically consistent, we will analyze the properties of the refraction index given in Eq.(\ref{IndiceRefraccionTOTAL}). Considering that $\epsilon(\omega)=n^{2}(\omega)$ is the permittivity of the material plates, we have:

\begin{equation}
\epsilon(\omega)-1=\frac{\omega_{\rm P}^{2}}{\omega_{0}^{2}-\omega^{2}-i\omega\widetilde{\gamma}(-i\omega)}.
\label{Permittivity}
\end{equation}
We can define the susceptibility kernel $\chi(\tau)$ for the model as \cite{Jackson}
\begin{equation}
\chi(\tau)=\frac{1}{2\pi}\int_{-\infty}^{+\infty}(\epsilon(\omega)-1)e^{-i\omega\tau}d\omega=\frac{\omega_{\rm P}^{2}}{2\pi}\int_{-\infty}^{+\infty}\frac{e^{-i\omega\tau}}{\omega_{0}^{2}-\omega^{2}-i\omega\widetilde{\gamma}(-i\omega)}d\omega.
\label{Susceptibility}
\end{equation}
In principle,  this integral can be evaluated by contour integration. Inversely, the permittivity can be expressed in terms of $\chi(\tau)$ as:
\begin{equation}
\epsilon(\omega)=1+\int_{-\infty}^{+\infty}\chi(\tau)e^{i\omega\tau}d\tau,
\label{Permi-Suscep}
\end{equation}
which can be viewed as a representation of $\epsilon(\omega)$ in the complex $\omega$-plane. The permittivity
is well defined when $\chi(\tau)$ is finite for all $\tau$ and  $\chi(\tau)\rightarrow 0$ as $\tau\rightarrow\pm\infty$,
and its analytical properties
can be studied directly from this expression.

All properties of the permittivity and susceptibility functions are strongly dependent on the Laplace transform of the damping kernel $\widetilde{\gamma}(ik)$, which in turn depends on the spectral density of the environment.
After taking the Laplace tranform of  Eq.(\ref{QBMNucleoAmortiguamiento}), we obtain

\begin{equation}
\widetilde{\gamma}(s)=\frac{2}{m}\int_{0}^{+\infty}d\omega\frac{J(\omega)}{\omega}\frac{s}{(s^{2}+\omega^{2})}.
\label{NucleoAmortiguamientoLaplace}
\end{equation}
As already described, a physical spectral density must incorporate a cutoff function. One could use a sharp cutoff or, alternatively,
choose a continuous cutoff function that approaches zero rapidly for frequencies greater than the cutoff frequency $\Lambda$, ensuring the convergence of the integral.

The first alternative, although simpler, makes the function $\widetilde{\gamma}(ik)$
not  well-defined in the complex plane. The second alternative solves this problem and allows the use of the residue theorem to evaluate the integral.
Inserting Eq.(\ref{QBMDensidadEspectralEntornos}) into Eq.(\ref{NucleoAmortiguamientoLaplace}), we obtain

\begin{equation}
\widetilde{\gamma}(s)=\frac{4\gamma_{0}s}{\pi\Lambda^{\alpha-1}}\int_{0}^{+\infty}d\omega\frac{\omega^{\alpha-1}}{(s^{2}+\omega^{2})}\textit{f}\left(\frac{\omega}{\Lambda}\right).
\label{NucleoAmortiguamientoLaplaceConJ}
\end{equation}

In order to apply  the residue theorem  the integrand must be holomorphic on the superior complex half-plane,  except in a finite number of points which are not on the real axis. Thus, different results are obtained considering distributions with or without poles on the superior half-plane. For an ohmic environment ($\alpha=1$) and no cutoff function it is easy to see that $\widetilde{\gamma}(s)=2\gamma_{0}$.

In the case of distributions without poles (for example, a gaussian cutoff function), we have:

\begin{equation}
\widetilde{\gamma}_{NP}(-ik)=\frac{\pi}{mk}J(k)\equiv\widetilde{\gamma}_{1}(k),
\label{GammaTransformadaIKSinPolos}
\end{equation}
where the subscript NP denotes the fact that the distribution has no poles. The resulting function is real and even in the variable $k$.

On the other hand, the distributions usually considered in the literature have poles on $\pm i\Lambda$ (for example, a Lorentzian distribution). In these cases, for odd $\omega^{\alpha}$ (with $\alpha<4$ to mantain the convergence in
Eq.(\ref{NucleoAmortiguamientoLaplaceConJ})), we get

\begin{equation}
\widetilde{\gamma}_{P}(-ik)=\frac{\pi}{mk}J_{\Lambda}(k)+i(-1)^{\frac{\alpha-1}{2}}\frac{\pi}{mk}\left(-\frac{k}{\Lambda}\right)^{\alpha-1}J_{-k}(\Lambda),
\label{GammaTransformadaIKConPolos}
\end{equation}
where the subscripts on $J$ denote the location of the pole. Although the resulting function is complex, the second equality in Eq.(\ref{GammaTransformadaIKSinPolos}) remains valid.

Taking into account the above properties of the damping kernel, we now continue analyzing the properties of the permittivity and susceptiblility functions. As a particular example, in the Drude model one has $\widetilde{\gamma}(-i\omega)\equiv\gamma_0$. Therefore the denominator in Eq.(\ref{Susceptibility}) has two poles, both in the lower half $\omega$-plane. Thus,
as expected from a physical point of view,
the  susceptibility kernel shows a causal behavior,  since it vanishes for $\tau<0$. The analyticity of $\epsilon(\omega)$ in the upper half $\omega$-plane allows the use of Cauchy's theorem, resulting in
the well-known Kramers-Kronig relations for the real and imaginary part of the permittivity function $\epsilon(\omega)$.

In our more general case, the physical properties of $\epsilon(\omega)$ are determined by  the function $\widetilde{\gamma}(-i\omega)$. This dissipation function is given by the theory of quantum open systems through Eqs.(\ref{GammaTransformadaIKSinPolos}) and (\ref{GammaTransformadaIKConPolos}), and depends on the chosen cutoff function.

Let us first consider, for simplicity,  the case in which the cutoff function has no poles, represented by the Eq.(\ref{GammaTransformadaIKSinPolos}).  For a given spectral density, the denominator in Eq.(\ref{Susceptibility})  reads
\begin{equation}
D_{NP}^{(\alpha)}(\omega)=\omega_{0}^{2}-\omega^{2}-i2\gamma_{0}\omega\left(\frac{\omega}{\Lambda}\right)^{\alpha-1}\textit{f}\left(\frac{\omega}{\Lambda}\right).
\label{DenominatorNP}
\end{equation}

If we choose an ohmic environment ($\alpha=1$) and no cutoff function (which is equivalent to put $\textit{f}\equiv 1$), we reobtain the Drude model (if $\omega_0=0$) or the one-resonance model (when $\omega_0 \not= 0$). In principle, we could consider
other values of $\alpha$ while  keeping  $\textit{f}\equiv 1$. In this case,   $\omega^{\alpha}$ should be an odd function. For example, $\alpha=3$ gives an ill-defined pole configuration,  since one of the poles lies on the upper half $\omega$-plane, breaking the analyticity of the integrand in Eq.(\ref{Susceptibility}), and resulting in a non-causal susceptibility, which turns out to be unphysical.

Therefore we see that for this supraohmic environment the use of a cutoff is unavoidable.
We may use an analytical cutoff (like a gaussian function), or a lorentzian cutoff function.
The first alternative leads to a denominator $D_{NP}^{(\alpha)}$ whose zeroes cannot be obtained analytically.
The second alternative, valid for $\alpha<4$ such that $\omega^{\alpha}$ is an odd function, leads to a denominator
\begin{equation}
D_{P}^{(\alpha)}(\omega)=\frac{(\Lambda^{2}+\omega^{2})(\omega_{0}^{2}-\omega^{2})+2\gamma_{0}\Lambda^{3-\alpha}\omega^{2}\left((-1)^{\frac{\alpha-1}{2}}\Lambda^{\alpha-2}-i\omega^{\alpha-2}\right)}{(\Lambda^{2}+\omega^{2})},
\label{DenominatorP}
\end{equation}
which for $\alpha=3$ gives
\begin{equation}
D_{P}^{(3)}(\omega)=\frac{\Lambda\omega_{0}^{2}-i\omega_{0}^{2}\omega-(2\gamma_{0}+\Lambda)\omega^{2}+i\omega^{3}}{\Lambda-i\omega}.
\label{DenominatorP3}
\end{equation}

We denote the zeroes of $D_{P}^{(3)}$ as $\omega_{i}=\omega_{0}x_{i}$ (with $i=1,2,3$). The three roots turn out to be located in the lower half $\omega$-plane, which ensures the causality property. Also, one of the roots is purely imaginary ($x_{1}=-x_{1}^{*}=-i|x_{1}|$) and the others two have the same (negative) imaginary part but opposite real parts ($x_{3}=-x_{2}^{*}$). Thus, the susceptibility kernel reads

\begin{equation}
\chi_{P}^{(3)}(\tau)=-\left(\frac{\omega_{P}}{\omega_{0}}\right)^{2}\left[\frac{(\Lambda-\omega_{0}|x_{1}|)}{(x_{1}-x_{2})(x_{1}+x_{2}^{*})}e^{-\omega_{0}|x_{1}|\tau}+2Re\left[\frac{(\Lambda-i\omega_{0}x_{2})}{(x_{2}-x_{1})(x_{2}+x_{2}^{*})}e^{-i\omega_{0}x_{2}\tau}\right]\right]\theta(\tau),
\label{SusceptibilityP3}
\end{equation}
where it is clear that it is a causal real function and, due to the negativity of the imaginary part of the roots $x_{i}$, we have $\chi_{P}^{(3)}(\tau)\rightarrow 0$ for $\tau\rightarrow+\infty$ as is expected.
We also have that $\chi_{P}^{(3)}(0)=0$ but $\chi_{P}^{(3)'}(0)\neq0$, and therefore the asymptotic expression found in \cite{Jackson} still remains valid, as well as the Kramers-Kronig relations.

It is worth noting that an ohmic environment ($\alpha=1$) can be also studied with a cutoff function, obtaining similar results.

All in all, we have shown that our model is physically consistent and generalizes previous results for the permittivity of absorbing media, including as particular case the Drude model.  Plasma-like models do not contain dissipation and can be obtained by taking $\gamma_0 =0$, that corresponds to no coupling between the system and the bath.

\section{Casimir Force}

\subsection{The Energy-Momentum Tensor and the Different Contributions to the Casimir Force}

Once determined the two contributions to the field, we proceed to compute the Casimir force between the plates as given by Eq. (\ref{FCasimirPresion}). For this purpose, it is necessary to compute the expectation value of $\widehat{T}_{xx}$ which is given by Eq. (\ref{T001D}).

Considering the vacuum and Langevin contributions according to Eq.(\ref{SeparacionContribuciones}), we have

\begin{eqnarray}
\widehat{T}_{xx}(x;t)&=&\frac{1}{2}\left((\partial_{0}(\widehat{\phi}_{V}+\widehat{\phi}_{L}))^2
+(\partial_{x}(\widehat{\phi}_{V}+\widehat{\phi}_{L}))^2\right)=\widehat{T}_{xx}^{V}\otimes\mathbb{I}_{S}
\otimes\mathbb{I}_{B}+\mathbb{I}_{\phi}\otimes\mathbb{I}_{S}\otimes\widehat{T}_{xx}^{L}+
\left(\partial_{x}\widehat{\phi}_{V}\right)\otimes\mathbb{I}_{S}\otimes\left(\partial_{x}\widehat{\phi}_{L}\right)\nonumber\\&&
+\left(\partial_{t}\widehat{\phi}_{V}\right)\otimes\mathbb{I}_{S}\otimes\left(\partial_{t}\widehat{\phi}_{L}\right).
\label{T001DTOTAL}
\end{eqnarray}
It is worth to remark  that there are cross terms which act over two parts of the total Hilbert space.

As we are interested in the steady state of the system, which is assumed at thermal equilibrium, each part of the total density matrix is represented by a thermal-type density matrix. On the other hand, both field contributions $\widehat{\phi}_{V}$ and $\widehat{\phi}_{L}$ are linear on the annihilation and creation operators of their respective parts of the total Hilbert space. Thus, the cross terms do not contribute to the force in the case of thermal equilibrium.
Then, the problem is reduced to compute the expectation values over thermal states of the operators $\widehat{T}_{xx}^{V}$ and $\widehat{T}_{xx}^{L}$, i.e.,

\begin{equation}
\mathfrak{f}=\left\langle\widehat{T}_{xx}\right\rangle=Tr_{\phi}\left(\widehat{\rho}_{\phi}\widehat{T}_{xx}^{V}\right)+Tr_{B}\left(\widehat{\rho}_{B}\widehat{T}_{xx}^{L}\right)=\mathfrak{f}_{V}+\mathfrak{f}_{L}.
\label{ExpectacionT001DTOTAL}
\end{equation}
Thus, the Casimir force also have two contributions:
\begin{equation}
F_{C}=\mathfrak{f}_{I}-\mathfrak{f}_{III}=\left(\mathfrak{f}_{I}^{V}+\mathfrak{f}_{I}^{L}\right)-\left(\mathfrak{f}_{III}^{V}+\mathfrak{f}_{III}^{L}\right)=F_{C}^{V}+F_{C}^{L}.
\label{FuerzaCasimirTOTAL}
\end{equation}

\subsection{Vacuum Casimir Force}

For the vacuum contribution, $\widehat{T}_{xx}^{V}$ is quadratic in the annihilation and creation operators. Thus, in order compute the expectation value over the thermal state, we need to evaluate the expectation values of the products of the annihilation and creation operators. These are given by the known expressions:
\begin{equation}
\left\langle\widehat{a}_{k}\widehat{a}_{k'}\right\rangle=\left\langle\widehat{a}_{k}^{\dag}\widehat{a}_{k'}^{\dag}\right\rangle=0,
\label{ExpectacionAA0}
\end{equation}
\begin{equation}
\left\langle\widehat{a}_{k}\widehat{a}_{k'}^{\dag}\right\rangle=\delta(k-k')(1+N(\omega_{k})),
\label{ExpectacionAAdDelta1masN}
\end{equation}
\begin{equation}
\left\langle\widehat{a}_{k}^{\dag}\widehat{a}_{k'}\right\rangle=\delta(k-k')N(\omega_{k}),
\label{ExpectacionAdADeltaN}
\end{equation}
where $N(\omega_{k})=\frac{1}{e^{\beta\omega_{k}}-1}$.

Taking into account Eq.(\ref{OperadorPhiDielectrico}),  we have, in region $l$
\begin{equation}
\mathfrak{f}_{l}^{V}(x)=Tr_{\phi}\left(\widehat{\rho}_{\phi}\widehat{T}_{xx}^{V,l}\right)=\frac{1}{4}\left[\int_{0}^{+\infty}+\int_{-\infty}^{0}\right]\frac{dk}{2\pi}\coth\left(\frac{\beta\omega_{k}}{2}\right)\left(\omega_{k}|f_{k}^{l}(x)|^{2}+\frac{1}{\omega_{k}}\left|\frac{df_{k}^{l}}{dx}\right|^{2}\right),
\label{DensEnLibreVacioTOTAL}
\end{equation}
which is identical to the expression for a non-absorbing medium except that, in this case, there is a thermal factor $\coth(\beta\omega_{k}/2)$ related to the temperature of the field.

Using the solutions for the modes functions $f_{k}^{l}$ in regions I and III
(see Appendix A), the vacuum Casimir force is given by
\begin{equation}
F_{C}^{V}=\mathfrak{f}_{I}^{V}-\mathfrak{f}_{III}^{V}=\frac{1}{2}\int_{0}^{+\infty}\frac{dk}{2\pi}k\coth\left(\frac{\beta \omega_k}{2}\right)\left[1+|R_{k}|^2+|T_{k}|^2-2\left(|C_{k}|^2+|D_{k}|^2\right)\right].
\label{FuerzaCasimirVacioTOTAL}
\end{equation} The coefficients $R_k$, $T_k$, $C_k$, and $D_k$ are explicitly given in Appendix A.
It is worth noting the appearance of a thermal global factor in the last expression, which comes from the field's thermal state at temperature $T$, based on the equilibrium assumption.

\subsection{Langevin Casimir Force}

For computing the Langevin contribution to the force, it is necessary to know the expectation value, over the bath's thermal state, of the force operator, where binary products are evaluated at different frequencies.

For any time-dependent hermitian operator, the expectation value evaluated at different times, corresponds to the correlation function of the operator. This matches with one-half of the anti-commutator expectation value at different times. Thus, making Fourier transforms over both times, we can compute the desired products  of the Fourier transform of the force operator, at different frequencies.

For the case of thermal equilibrium, the anti-commutator expectation value at different times of the force operator is provided by the QBM theory. One can show that it matches  the noise kernel $D_{1}(t-t')$ of Eq.(\ref{QBMNucleoRuido}). Thus, we obtain
\begin{equation}
\left\langle\left\{\widehat{F}(k);\widehat{F}(k')\right\}\right\rangle=J(\omega_k)\coth\left(\frac{\beta\omega_k}{2}\right)\delta(k+k').
\label{ExpectacionAntiFourierOperadorF}
\end{equation}

Considering that in our case the stochastic force operator depends on the position where the atom is located, we finally have
\begin{equation}
\left\langle\widehat{F}(x;k)\widehat{F}(x';k')\right\rangle=\delta(x-x')\frac{J(\omega_k)}{2\eta}\coth\left(\frac{\beta\omega_k}{2}\right)\delta(k+k'),
\label{ExpectacionProdBinFourierOperadorFTOTAL}
\end{equation}
where we have included the atom density $\eta$ due to dimensional issues. As it can be seen, the  frequency spectrum is not flat.

Taking all this into account, $\widehat{T}_{xx}^L$ in regions I and III are given by:

\begin{equation}
\widehat{T}_{xx}^{L,I}(x;t)=\int_{-\infty}^{+\infty}\frac{dk}{2\pi}\int_{-\infty}^{+\infty}\frac{dk'}{2\pi}(-kk')\widehat{W}_{1}(k)\widehat{W}_{1}(k')e^{-i(k+k')(x+t)},
\label{T001DRegILangevinTOTAL}
\end{equation}

\begin{equation}
\widehat{T}_{xx}^{L,III}(x;t)=\int_{-\infty}^{+\infty}\frac{dk}{2\pi}\int_{-\infty}^{+\infty}\frac{dk'}{2\pi}(-kk')e^{-i(k+k')t}\left(\widehat{W}_{2}(k)\widehat{W}_{2}(k')e^{i(k+k')x}+\widehat{W}_{3}(k)\widehat{W}_{3}(k')e^{-i(k+k')x}\right).
\label{T001DRegIIILangevinTOTAL}
\end{equation}
The  coefficients $\widehat W_i(k)$ in these equations are given in Appendix B, and are  linear functions of the Fourier
transform of the stochastic force operator. Taking into account the explicit expressions in Appendix B and Eq.(\ref{ExpectacionProdBinFourierOperadorFTOTAL}), the desired expectation values are:
\begin{eqnarray}
\left\langle\widehat{W}_{1}(k)\widehat{W}_{1}(k')\right\rangle&=&|\mathfrak{W}(k)|^{2}\frac{2\pi}{m}\frac{J(\omega_k)}{k^{2}\widetilde{\gamma}_{1}(k)}e^{-z_{2}}\coth\left(\frac{\beta\omega_k}{2}\right)\delta(k+k')\nonumber\\&\times &
\Big[n_{1}\left(1-e^{-z_{2}}\right)\left(|t|^{2}e^{z_{2}}+|r_{n}+re^{i2ka}|^{2}+e^{z_{2}}|1+rr_{n}e^{i2ka}|^{2}+|t|^{2}|r_{n}|^{2}\right)
\nonumber \\ &+& 2n_{2}Re\left(|t|^{2}i\left(e^{-iz_{1}}-1\right)
r_{n}^{\ast}+i\left(1-e^{iz_{1}}\right)\left(1+r^{\ast}r_{n}^{\ast}e^{-i2ka}\right)\left(r_{n}+re^{i2ka}\right)\right)\Big],
\label{ExpectacionW1W1TOTAL}
\end{eqnarray}
\begin{eqnarray}
\left\langle\widehat{W}_{2}(k)\widehat{W}_{2}(k')\right\rangle&=&\left\langle\widehat{W}_{3}(k)\widehat{W}_{3}(k')\right\rangle=|\mathfrak{W}(k)|^{2}\frac{2\pi}{m}\frac{J(\omega_k)}{k^{2}\widetilde{\gamma}_{1}(k)}\left(1+|r|^{2}\right)
\coth\left(\frac{\beta\omega_k}{2}\right)\delta(k+k')\nonumber \\ &\times & \Big[n_{1}\left(1-e^{-z_{2}}\right)\left(1+|r_{n}|^{2}e^{-z_{2}}\right)+2n_{2}e^{-z_{2}}Re\left(i\left(1-e^{iz_{1}}\right)r_{n}\right)\Big],
\label{ExpectacionW2W2oW3W3TOTAL}
\end{eqnarray}

\noindent where $n=n_{1}+in_{2}$, i.e.,
$n_{1}={\rm Re}(n)$ and $n_{2}={\rm Im}(n)$, $\widetilde{\gamma}(ik)=\widetilde{\gamma}_{1}(k)+i\widetilde{\gamma}_{2}(k)$,  $z_{1}=2kn_{1}d$ and
$z_{2}=2kn_{2}d$. The explicit expressions for the coefficients $r,r_n$ and $t$ can be found in Appendix A.

Therefore, the Langevin contribution to the force in regions I and III are given by
\begin{eqnarray}
\mathfrak{f}_{I}^{L}&=&Tr_{B}\left(\widehat{\rho}_{B}\widehat{T}_{xx}^{L,I}\right)=\int_{-\infty}^{+\infty}\frac{dk}{2\pi}~|\mathfrak{W}(k)|^{2}~\frac{2\pi}{m}\frac{J(\omega_k)}{\widetilde{\gamma}_{1}(k)}~e^{-z_{2}}\coth\left(\frac{\beta\omega_k}{2}\right)\nonumber \\ &\times & \Big[n_{1}\left(1-e^{-z_{2}}\right)\left(|t|^{2}e^{z_{2}}+|r_{n}+re^{i2ka}|^{2}+e^{z_{2}}|1+rr_{n}e^{i2ka}|^{2}+|t|^{2}|r_{n}|^{2}\right)\nonumber \\ &+& 2n_{2}Re\left(|t|^{2}i\left(e^{-iz_{1}}-1\right)r_{n}^{\ast}+i\left(1-e^{iz_{1}}\right)\left(1+r^{\ast}r_{n}^{\ast}e^{-i2ka}\right)\left(r_{n}+re^{i2ka}\right)\right)\Big],
\label{DensEnLibreLangevinRegITOTALSinParidad}
\end{eqnarray}
\begin{eqnarray}
\mathfrak{f}_{III}^{L}&=&Tr_{B}\left(\widehat{\rho}_{B}\widehat{T}_{xx}^{L,III}\right)=\int_{-\infty}^{+\infty}\frac{dk}{2\pi} ~|\mathfrak{W}(k)|^{2}~\frac{4\pi}{m}\frac{J(\omega_k)}{\widetilde{\gamma}_{1}(k)}~\coth\left(\frac{\beta\omega_k}{2}\right)\left(1+|r|^{2}\right)\nonumber\\ &\times & \Big[n_{1}\left(1-e^{-z_{2}}\right)\left(1+|r_{n}|^{2}e^{-z_{2}}\right)+2n_{2}e^{-z_{2}}Re\left(i\left(1-e^{iz_{1}}\right)r_{n}\right)\Big].
\label{DensEnLibreLangevinRegIIITOTALSinParidad}
\end{eqnarray}

Taking advantage that the integration is over all the values of $k$, the fact that the change $k\leftrightarrow -k$ is equivalent to complex conjugation, and the second equality of Eq. (\ref{GammaTransformadaIKSinPolos}), we obtain
\begin{equation}
\mathfrak{f}_{I}^{L}=\int_{0}^{+\infty}\frac{dk}{2\pi}\frac{k}{2}\frac{|n+1|^{2}}{|n|^{2}}\frac{8|t|^{2}n_{1}e^{-z_2}}{|1-r^{2}e^{i2ka}|^{2}}\coth\left(\frac{\beta
k}{2}\right)\left(1-e^{-z_{2}}\right)\left(|t|^{2}e^{z_{2}}+|r_{n}+re^{i2ka}|^{2}+e^{z_{2}}|1+rr_{n}e^{i2ka}|^{2}+|t|^{2}|r_{n}|^{2}\right),
\label{DensEnLibreLangevinRegITOTAL}
\end{equation}

\begin{equation}
\mathfrak{f}_{III}^{L}=\int_{0}^{+\infty}\frac{dk}{2\pi}\frac{k}{2}\frac{|n+1|^{2}}{|n|^{2}}\frac{16|t|^{2}n_{1}}{|1-r^{2}e^{i2ka}|^{2}}\coth\left(\frac{\beta
k}{2}\right)\left(1-e^{-z_{2}}\right)\left(1+|r|^{2}\right)\left(1+|r_{n}|^{2}e^{-z_{2}}\right).
\label{DensEnLibreLangevinRegIIITOTAL}
\end{equation}

Note that the presence of the thermal factor $\coth\left(\frac{\beta k}{2}\right)$ is in agreement with the null temperature result obtained in other works for an ohmic environment \cite{Dorota} since when $T\rightarrow 0$, $\coth\left(\frac{\beta k}{2}\right)\rightarrow 1$.

Finally, the Langevin contribution for the Casimir force is:

\begin{eqnarray}
F_{\rm C}^{L}&=&\mathfrak{f}_{I}^{L}-\mathfrak{f}_{III}^{L}=\int_{0}^{+\infty}\frac{dk}{2\pi}\frac{k}{2}\frac{|n+1|^{2}}{|n|^{2}}\frac{8|t|^{2}n_{1}}{|1-r^{2}e^{i2ka}|^{2}}\coth\left(\frac{\beta
k}{2}\right)\nonumber\\&\times &\left(1-e^{-z_{2}}\right)\left(|t|^{2}+|r_{n}+re^{i2ka}|^{2}e^{-z_{2}}+|1+rr_{n}e^{i2ka}|^{2}+|t|^{2}|r_{n}|^{2} e^{-z_{2}}-2\left(1+|r|^{2}\right)\left(1+|r_{n}|^{2}e^{-z_{2}}\right)\right).
\label{FuerzaCasimirLangevinTOTAL}
\end{eqnarray}

It should be noted that here also appears a global thermal factor, as in the vacuum case, but this comes from the bath's temperature while in the vacuum case comes from the field's equilibrium temperature.  Note also that $F_{\rm C}^L$ vanishes when there is no coupling to an
environment, since in this case the refraction index is real and
therefore $z_2=0$.

\subsection{Total Casimir Force}

The total Casimir force is determined from the expressions
Eq. (\ref{FuerzaCasimirTOTAL}), (\ref{FuerzaCasimirVacioTOTAL}) and (\ref{FuerzaCasimirLangevinTOTAL}). The resulting force can be written in a very compact form. On the one hand, it can be proved that the total free energy in region I (outside the plates) coincides with that for free field at temperature $T$ i.e.,
\begin{equation}
\mathfrak{f}_{I}=\mathfrak{f}_{I}^{V}+\mathfrak{f}_{I}^{L}=\int_{0}^{+\infty}\frac{dk}{2\pi}k\coth\left(\frac{\beta
k}{2}\right), \label{DensEnLibreRegITOTAL}
\end{equation}
 which is expectable due to translational invariance outside the plates and our assumption of thermal equilibrium. On the other hand, in region III we have
\begin{equation}
\mathfrak{f}_{III}=\mathfrak{f}_{III}^{V}+\mathfrak{f}_{III}^{L}=\int_{0}^{+\infty}\frac{dk}{2\pi}k\coth\left(\frac{\beta
k}{2}\right)\frac{(1-|r|^{4})}{|1-r^{2}e^{i2ka}|^{2}}.
\label{DensEnLibreRegIIITOTAL}
\end{equation}
Therefore, the total Casimir force is finally written as:
\begin{equation}
F_{\rm C}[a]=\mathfrak{f}_{I}-\mathfrak{f}_{III}=\int_{0}^{+\infty}\frac{dk}{2\pi}k\coth\left(\frac{\beta
k}{2}\right)\left(1-\frac{1-|r|^{4}}{|1-r^{2}e^{i2ka}|^{2}}\right) \, ,
\label{FuerzaCasimirTOTALFinal}
\end{equation}
or equivalently
\begin{equation}
F_{\rm C}[a]=-\frac{1}{\pi}Re\Big[\int_{0}^{+\infty}dk \, k\coth\left(\frac{\beta
k}{2}\right) \frac{r^2 e^{i2ka}}{1-r^{2}e^{i2ka}}\Big] \, .
\label{FuerzaCasimirTOTALFinal2}
\end{equation}
This expression is formally identical to the case of dissipative material (ohmic environment) at zero temperature found in previous works \cite{Dorota, Dorota90}. However, it  contains two generalizations: on the one hand, temperature has been included in the formalism in a natural way by means of the open quantum system theory. On the other hand,
the equation is valid for a general environment:  the  refraction index is in general complex,  and dependent on the function $\widetilde{\gamma}(ik)$, which comes from the interactions at the microscopic level.
The associated permittivity depends on the type of environment, and reproduces known results (as the Drude model)
as particular examples.

\subsection{Convergence and Limits}

In this Section we will study some properties and limits of our final result  Eq.(\ref{FuerzaCasimirTOTALFinal2}).
Let us first study the convergence of this expression. In general, the Casimir force calculations involve several regularization methods to achieve a finite result.  A usual approach is to introduce a high frequency cutoff,  in order to take into account that
real materials are transparent at high frequencies. This characteristic is already incorporated in our model. Indeed,
the complex refraction index $n$  includes all the environment properties which produce dissipation and noise. For large values of $k$, taking into account   Eq.(\ref{IndiceRefraccionTOTAL}) one can check that

\begin{equation}
n_{1}\rightarrow 1-\frac{2\pi\eta e^{2}}{m}\frac{1}{k^{2}},
\label{LimiteKGrandeParteRealIndiceRefraccion}
\end{equation}

\begin{equation}
n_{2}\rightarrow \frac{2\pi\eta
e^{2}}{m}\frac{\widetilde{\gamma}_{1}(k)}{k^{3}}.
\label{LimiteKGrandeParteImaginariaIndiceRefraccion}
\end{equation}
Then, in the same limit the reflection coefficient $r$ behaves as

\begin{equation}
r\rightarrow \frac{\pi\eta
e^{2}}{m}\frac{1}{k^{2}}\left(1-e^{i2kd}\right) \, ,
\label{LimiteKGrandeR}
\end{equation}
and in consequence the integrand of Eq.(\ref{FuerzaCasimirTOTALFinal2}) is $O(k^{-3})$ for large values of $k$.
Thus the convergence is ensured when $k\rightarrow +\infty$, regardless the temperature and the type of environment
considered.

On the other hand, Eq.(\ref{FuerzaCasimirTOTALFinal2}) contains as particular cases some known results.
The non-absorbing medium case can be easily obtained by setting the relaxation constant $\gamma_{0} = 0$ in all the expressions. This makes the refraction index real, which cancel the Langevin contribution in Eq.(\ref{FuerzaCasimirLangevinTOTAL}) since the factor $1-e^{-z_{2}} \rightarrow 0$. Thus, only the vacuum contribution in Eq.(\ref{FuerzaCasimirVacioTOTAL}) survives but with a real refraction index \cite{Dorota90}.

Another important limit is the well-known Lifshitz formula. In the original work \cite{Lifshitz}, Lifshitz considered
semispaces separated by a finite distance.  Therefore  one should take the limit of large thickness ($d\rightarrow +\infty$),
in which $r$ should be replaced by $r_{n}$. After a rotation to the imaginary-frequency  axis Eq.(\ref{FuerzaCasimirTOTALFinal2}) becomes, in the $T=0$ case
\begin{equation}
F_{\rm C}[a]=\frac{1}{\pi}\int_{0}^{+\infty}dss\frac{r_{n}^{2}(is)e^{-2sa}}{1-r_{n}^{2}(is)e^{-2sa}},
\label{FuerzaCasimirFormulaLifshitz}
\end{equation}
which is of the form of Lifshitz formula for a scalar field in $1+1$ dimensions.  In the case $T\neq 0$,  we must take into account that
$\coth\left(\frac{\beta
k}{2}\right)$ has poles on the imaginary axis at the Matsubara frequencies  $2\pi i \beta j=i\xi_j, j=0,1,2,...$. Therefore, the integration path
can be rotated to the imaginary axis, but must be deformed to avoid the poles. This is a standard procedure, that converts the integral over frequencies into the Matsubara sum
\begin{equation}
F_{\rm C}[a]=2T\sum_{j\geq 1}\xi_j\frac{r_{n}^{2}(i\xi_j)e^{-2\xi_ja}}{1-r_{n}^{2}(i\xi_j)e^{-2\xi_ja}},
\label{Matsubara}
\end{equation}
which is the standard expression for Lifshitz formula at $T\neq 0$.

It is interesting to remark that in our simplified $1+1$ model there is no discontinuity in the transition between the Drude and plasma
models. The Drude model is recovered, in the ohmic environment case, by setting $\omega_0=0$  
(i.e. free particles instead of harmonic oscillators for modelling the dielectric atoms) and, once this limit is taken, 
the plasma model corresponds to the particular case $\gamma_0=0$ (no coupling to the environment).  The reflection coefficient 
$r_n\to 1$ in the zero frequency limit, for any value of $\omega_{\rm P}$ and $\gamma_0$, even setting
$\gamma_0=0$ from the beginning. In the case of absence of coupling to an environment, of course one must assume
thermal equilibrium. Analogies between thermodynamics of a free Brownian particle and that of an electromagnetic field between 
two mirrors of finite conductivity have been studied in Ref. \cite{ingold}. 

\section{Connection with the Euclidean formalism}

Given that we are assuming thermal equilibrium between the different parts of the system, the results presented in this paper
for the Casimir force could be derived following a functional approach in Euclidean space, as described for instance in Refs. \cite{plb08,Ccappa}.  We mention briefly the relation
between both approaches.

As is well known, in $1+1$ dimensions the free energy ${\mathcal E}$ for a quantum system in thermal equilibrium
at temperature $T$  can be computed as
\begin{equation}\label{eq:defe}
{\mathcal E} \;= - T
\log \frac{\mathcal Z(a)}{{\mathcal Z}(\infty )} \;,
\end{equation}
where ${\mathcal Z}(a)$ is the partition function when the plates are separated by a distance $a$.
The partition function can be represented by the functional integral
\begin{equation}
 {\mathcal Z} \,=\, \int {\mathcal D}\phi {\mathcal D}r {\mathcal D}q_n     \; e^{- S_E} \;,
\end{equation}
where $S_E$ is the Euclidean action for the full system. The integration is performed by imposing periodic boundary conditions
on the temporal coordinate.

After integrating the matter and bath degrees of freedom it is possible to find an effective action for the scalar field of the form
\begin{equation}
S_{\rm eff}=\int d^2x \frac{1}{2}\partial_\mu\phi\partial_\mu\phi +\int d^2x\int d^2x' V(x,x')\phi(x)\phi(x')\, ,
\end{equation}
and the vacuum persistence amplitude reads
\begin{equation}
 {\mathcal Z} \,=\, \int {\mathcal D}\phi     \; e^{- S_{\rm eff}} \;.
\end{equation}
The effective action for the scalar field is
quadratic because of the linear coupling we are choosing for the interaction between the vacuum field and the
matter degrees of freedom.  The potential $V(x,x')$ is different from zero only inside the plates,  and as in our model
the field $\phi(x,t)$ interacts only with the atom at $x$, it can be shown that the potential is of the form $V(x,t,x',t�)=\delta(x-x')
\lambda(t-t')$. The function $\lambda(t-t')$ encodes the information about the interaction of the vacuum field with the matter
degrees of freedom, and also of the influence of the thermal bath. Its Fourier transform is related to the
reflection coefficient of the slab.

Formally, the vacuum persistence amplitude is given by the functional determinant
\begin{equation}
{\mathcal Z} \,=\, \big(\det [-\Box +V] \big)^{-\frac{1}{2}} \;.
\end{equation}
An explicit evaluation of this  determinant leads to Lifshitz formula \cite{Ccappa}. So when considering thermal equilibrium,
one has an alternative route to the evaluation of the Casimir force, even when the field is coupled to other degrees of freedom.
However, this Euclidean functional approach would not be adequate  to compute the force for more other initial states
or, in general, in nonequilibrium situations. In that cases, the use of the theory of quantum open systems as described
in this paper is unavoidable.

\section{Conclusions}

In this paper we have presented a derivation of the Casimir force between two absorbing slabs in the framework of the theory of quantum open systems. We worked with a simplified model of a scalar vacuum field in $1+1$ dimensions. In order to describe the interaction of the vacuum field with the mirrors, we considered a model analogous to the HB model for QED, where the matter degrees of freedom are described by a continuous set of harmonic oscillators,  which are coupled not only to the vacuum field but also to a thermal bath that accounts for dissipative effects.

Following a standard procedure in the theory of quantum open systems, we showed that the field operator satisfies the modified Klein Gordon equation (\ref{EcMovCampoTOTAL}). This is a nonlocal Langevin equation, which describes the interaction
of the vacuum field with the matter degrees of freedom and the effects of the thermal bath on its dynamics (the environment is indirectly coupled to the quantum field through the matter). This equation is similar to the one that describes a Brownian particle coupled to an environment. Both the noise and dissipation
are determined by the properties of the environment.

The field operator that solves this "Klein Gordon- Langevin" equation can be written as the sum of two terms,  a vacuum contribution and a Langevin contribution. The same happens with the associated energy-momentum tensor, and therefore
we have a similar decomposition for the Casimir force between slabs. The final result for the total Casimir force is equivalent to
a $1+1$-version of the Lifshitz formula, expressed in terms of the reflection coefficients associated to the slabs.  Therefore, we
have presented, in this simplified model,  a first-principles derivation of Lifshitz formula in the framework of quantum
open systems.

The present work is closely related to Ref. \cite{Dorota}, that has been improved and generalized in several directions.
Indeed, that work assumes
the simplest forms for noise and dissipation (ohmic environment and constant dissipation), without specifying the properties of the environment.  Moreover it is doubtful whether a general non-ohmic environment can produce such effects at $T=0$. Here we
worked at $T\neq 0$,  and considered very general environments. We also linked the properties of the environment with
the macroscopic electromagnetic properties of the mirrors. There is also a close relation with the recent work \cite{daRosaetal3},
where the authors computed the force density associated to spatial variations of the permittivity.  As compared
with ours, in this reference the authors considered
the more realistic case of $3+1$ electromagnetic field, but only in the particular case of $T=0$ and constant dissipation. Moreover,
they did not consider the presence of boundaries as we did here, which allowed us to compute explicitly  the Casimir force between slabs and to obtain Lifshitz formula.

In order to apply the quantum open systems approach to a realistic calculation of the Casimir force,  we should generalize
our results to a $3+1$ model with  the electromagnetic field. Although technically more complex, we do not expect
conceptual complications in doing so. Regarding the long standing controversy about the temperature corrections
to the Casimir force, a crucial point is the behavior of the quantity 
\begin{equation}
\lim_{\zeta\to 0}\zeta^2[\epsilon(i\zeta)-1]\nonumber ,
\end{equation} 
which vanishes for the Drude model and is different from zero for the plasma model, producing in the last case an additional contribution
to the force coming from the TE zero mode. In the kind of microscopic models considered here, the TE zero mode
is suppressed as long as $\omega_0\neq 0$, as can be seen from Eq.(\ref{Permittivity}). 

Finally, in the $3+1$ dimensional models one could consider more general initial states and/or
non-equilibrium situations. We hope to address this issue in a forthcoming publication.

\section*{Acknowledgements}
This work was supported by UBA, CONICET and ANPCyT.

\appendix

\section{Boundary conditions and solutions for the vacuum contribution}
\label{vacuum}

In this Appendix we present the explicit form of the solutions of Eq. (\ref{EcModosVacio1DTOTAL}). En each region we have:

\begin{equation}
f_{k}^{I}(x)=e^{ikx}+R_{k}e^{-ikx}, \label{PropuestaFI+}
\end{equation}
\begin{equation}
f_{k}^{II}(x)=A_{k}e^{iknx}+B_{k}e^{-iknx}, \label{PropuestaFII+}
\end{equation}
\begin{equation}
f_{k}^{III}(x)=C_{k}e^{ikx}+D_{k}e^{-ikx}, \label{PropuestaFIII+}
\end{equation}
\begin{equation}
f_{k}^{IV}(x)=E_{k}e^{iknx}+F_{k}e^{-iknx}, \label{PropuestaFIV+}
\end{equation}
\begin{equation}
f_{k}^{V}(x)=T_{k}e^{ikx}. \label{PropuestaFV+}
\end{equation}

\noindent The different coefficients can be obtained by  imposing continuity of the mode functions and their derivatives
at the interfaces. They are given by:

\begin{equation}
R_{k}=e^{-ika}\left(r+\frac{t^{2}re^{i2ka}}{(1-r^{2}e^{i2ka})}\right),
\label{CoeficienteR}
\end{equation}
\begin{equation}
A_{k}=e^{-ik\frac{a}{2}}e^{ikn\frac{a}{2}}\frac{t(1+r_{n}re^{i2ka})}{t_{n}(1-r^{2}e^{i2ka})},
\label{CoeficienteA}
\end{equation}
\begin{equation}
B_{k}=e^{-ik\frac{a}{2}}e^{ikn\frac{a}{2}}\frac{t(r_{n}+re^{i2ka})}{t_{n}(1-r^{2}e^{i2ka})},
\label{CoeficienteB}
\end{equation}
\begin{equation}
C_{k}=\frac{t}{(1-r^{2}e^{i2ka})}, \label{CoeficienteC}
\end{equation}
\begin{equation}
D_{k}=\frac{tre^{ika}}{(1-r^{2}e^{i2ka})}, \label{CoeficienteD}
\end{equation}
\begin{equation}
E_{k}=\frac{t^{2}}{t_{n}(1-r^{2}e^{i2ka})}e^{ik\frac{a}{2}}e^{ikd}e^{-ikn\frac{a}{2}}e^{-iknd},
\label{CoeficienteE}
\end{equation}
\begin{equation}
F_{k}=\frac{t^{2}r_{n}}{t_{n}(1-r^{2}e^{i2ka})}e^{ik\frac{a}{2}}e^{ikd}e^{ikn\frac{a}{2}}e^{iknd},
\label{CoeficienteF}
\end{equation}
\begin{equation} T_{k}=\frac{t^2}{(1-r^{2}e^{i2ka})},
\label{CoeficienteT}
\end{equation}
for $k>0$ (while for $k<0$ the order of the solutions must be reversed and the refractive index and the coefficients should be conjugated), where $r=\frac{r_{n}(e^{i2knd}-1)}{(1-r_{n}^{2}e^{i2knd})}$ and
$t=\frac{4n}{(n+1)^2}\frac{e^{iknd}e^{-ikd}}{(1-r_{n}^{2}e^{i2knd})}$ are the reflection and transmission coefficients for one plate, while $r_{n}=\frac{n-1}{n+1}$ and $t_{n}=\frac{2n}{n+1}$ are the ones for an interface. The coefficients $R_{k}$ and $T_{k}$ can be interpreted as the reflection and transmission coefficients of the two plates configuration. However, it should be noted that, due to the presence of absorption, $|r|^{2}+|t|^{2}\neq 1$ and $|R_{k}|^{2}+|T_{k}|^{2}\neq 1$ in this case.

\section{Boundary conditions and solutions for the Langevin contribution}
\label{langevin}

In this Appendix we solve the equations  (\ref{EcMovCampoFourierLangevin1DReg135TOTAL}) and
(\ref{EcMovCampoFourierLangevin1DReg24TOTAL}). Due to the presence of a source in regions II and IV
(see Eq. (\ref{EcMovCampoFourierLangevin1DReg24TOTAL})), the solutions will have to parts: one associated to the homogeneous equation and other related directly to the source. Therefore, the solutions are:

\begin{equation}
\widehat{\phi}_{L}^{I}(x;k)=\widehat{W}_{1}(k)e^{-ikx},
\label{PropuestaLangevinI}
\end{equation}
\begin{equation}
\widehat{\phi}_{L}^{II}(x;k)=\widehat{U}_{1}(k)e^{iknx}+\widehat{U}_{2}(k)e^{-iknx}+\widehat{A}_{1}(x;k)e^{iknx}+\widehat{A}_{2}(x;k)e^{-iknx},
\label{PropuestaLangevinII}
\end{equation}
\begin{equation}
\widehat{\phi}_{L}^{III}(x;k)=\widehat{W}_{2}(k)e^{ikx}+\widehat{W}_{3}(k)e^{-ikx},
\label{PropuestaLangevinIII}
\end{equation}
\begin{equation}
\widehat{\phi}_{L}^{IV}(x;k)=\widehat{V}_{1}(k)e^{iknx}+\widehat{V}_{2}(k)e^{-iknx}+\widehat{B}_{1}(x;k)e^{iknx}+\widehat{B}_{2}(x;k)e^{-iknx},
\label{PropuestaLangevinIV}
\end{equation}
\begin{equation}
\widehat{\phi}_{L}^{V}(x;k)=\widehat{W}_{4}(k)e^{ikx},
\label{PropuestaLangevinV}
\end{equation}
with,

\begin{equation}
\widehat{A}_{1}(x;k)=\frac{1}{2n}\int_{-d-\frac{a}{2}}^{x}\widehat{G}(x';k)e^{-iknx'}dx',
\label{ParticularLangevinA1}
\end{equation}
\begin{equation}
\widehat{A}_{2}(x;k)=-\frac{1}{2n}\int_{-d-\frac{a}{2}}^{x}\widehat{G}(x';k)e^{iknx'}dx',
\label{ParticularLangevinA11}
\end{equation}
\begin{equation}
\widehat{B}_{1}(x;k)=\frac{1}{2n}\int_{\frac{a}{2}}^{x}\widehat{G}(x';k)e^{-iknx'}dx',
\label{ParticularLangevinB1}
\end{equation}
\begin{equation}
\widehat{B}_{2}(x;k)=-\frac{1}{2n}\int_{\frac{a}{2}}^{x}\widehat{G}(x';k)e^{-iknx'}dx',
\label{ParticularLangevinA111}
\end{equation}
where, for simplicity, we write $\widehat{G}(x;k)=\frac{4\pi\eta
e}{m}\frac{\widehat{F}(x;k)}{\left(k_{0}^{2}-k^{2}-ik\widetilde{\gamma}(-ik)\right)}$.

The coefficients $\widehat{W}_{l}(k)$, $\widehat{U}_{l}(k)$ and $\widehat{V}_{l}(k)$ are obtained by means of the appropriate boundary conditions. Thus, they are given by:

\begin{equation}
\widehat{W}_{1}(k)=\mathfrak{W}(k)e^{iknd}e^{-ik(a+d)}\left(\widehat{K}\left(1+rr_{n}e^{i2ka}\right)+\widehat{L}\left(r_{n}+re^{i2ka}\right) +\widehat{M}te^{ikd}e^{ik(a-nd)}+\widehat{N}tr_{n}e^{ikd}e^{ik(a+nd)}\right),
\label{CoeficienteOperadorW1}
\end{equation}

\begin{equation}
\widehat{W}_{2}(k)=\mathfrak{W}(k)\left(\widehat{K}r_{n}e^{i2knd}+\widehat{L}+\widehat{M}re^{ika}+\widehat{N}rr_{n}e^{ik(a+2nd)}\right),
\label{CoeficienteOperadorW2}
\end{equation}
\begin{equation}
\widehat{W}_{3}(k)=\mathfrak{W}(k)\left(\widehat{K}rr_{n}e^{ik(a+2nd)}+\widehat{L}re^{ika}+\widehat{M}+\widehat{N}r_{n}e^{i2knd}\right),
\label{CoeficienteOperadorW3}
\end{equation}
\begin{equation}
\widehat{U}_{1}=\frac{r_{n}}{t_{n}}e^{ik(n+1)(\frac{a}{2}+d)}\widehat{W}_{1},
\label{CoeficienteOperadorU1}
\end{equation}
\begin{equation}
\widehat{U}_{2}=\frac{1}{t_{n}}e^{ik(1-n)(\frac{a}{2}+d)}\widehat{W}_{1},
\label{CoeficienteOperadorU2}
\end{equation}
\begin{equation}
\widehat{V}_{1}=\frac{1}{t_{n}}e^{ik(1-n)\frac{a}{2}}(\widehat{W}_{2}+r_{n}e^{-ika}\widehat{W}_{3}),
\label{CoeficienteOperadorV1}
\end{equation}
\begin{equation}
\widehat{V}_{2}=\frac{1}{t_{n}}e^{ik(n-1)\frac{a}{2}}(r_{n}e^{ika}\widehat{W}_{2}+\widehat{W}_{3}),
\label{CoeficienteOperadorV2}
\end{equation}
\begin{equation}
\widehat{W}_{4}=e^{ik(n-1)d}(\widehat{W}_{2}+r_{n}e^{-ika}\widehat{W}_{3}+\frac{e^{-ik\frac{a}{2}}}{(n+1)}\widehat{N}),
\label{CoeficienteOperadorW4}
\end{equation}

\noindent where,

\begin{equation}
\widehat{K}=e^{ikn\frac{a}{2}}\int_{-d-\frac{a}{2}}^{-\frac{a}{2}}\widehat{G}(x;k)e^{iknx}dx,
\label{IntegralLangevinK}
\end{equation}

\begin{equation}
\widehat{L}=e^{-ikn\frac{a}{2}}\int_{-d-\frac{a}{2}}^{-\frac{a}{2}}\widehat{G}(x;k)e^{-iknx}dx,
\label{IntegralLangevinL}
\end{equation}

\begin{equation}
\widehat{M}=e^{-ikn\frac{a}{2}}\int_{\frac{a}{2}}^{d+\frac{a}{2}}\widehat{G}(x;k)e^{iknx}dx,
\label{IntegralLangevinM}
\end{equation}

\begin{equation}
\widehat{N}=e^{ikn\frac{a}{2}}\int_{\frac{a}{2}}^{d+\frac{a}{2}}\widehat{G}(x;k)e^{-iknx}dx,
\label{IntegralLangevinN}
\end{equation}

\begin{equation}
\mathfrak{W}(k)=\frac{we^{ik\frac{a}{2}}}{2n\left(1-r^{2}e^{i2ka}\right)},\;\;\;\;\textit{with}\;\;\;\;w=\frac{2n}{(n+1)\left(1-r_{n}^{2}e^{i2knd}\right)}.
\label{WdeKyw}
\end{equation}

The Langevin contribution is evaluated in the five regions. Since $\widehat{K},\widehat{L},\widehat{M}$ and $\widehat{N}$ depend linearly on the Fourier transform of the stochastic force operator, it should be noted that the coefficients also depend on the same way. In fact, since the stochastic force operator depend linearly on the bath's annihilation and creation operators, the Langevin contribution depend in that way too.

\end{document}